\documentclass[aip,endfloats,unsortedaddress,preprint]{revtex4-1}
\usepackage{graphicx}
\usepackage{color}
\usepackage{epsfig}
\usepackage{dcolumn}
\usepackage{ulem}

\begin{document}
\definecolor{red}{rgb}{1,0,0}
\title{Thermodynamics of Water Entry in Hydrophobic Channels of Carbon Nanotubes}
\author{Hemant Kumar}
\author{Biswaroop Mukherjee}
\affiliation{Centre For Condensed Matter Theory, Department of Physics, Indian Institute of Science, Bangalore, 560 012, India}
\author{Shiang-Tai Lin}
\affiliation{Department of Chemical Engineering, National Taiwan University, Taipei, 10617, Taiwan}
\author{Chandan Dasgupta}
\affiliation{Centre For Condensed Matter Theory, Department of Physics, Indian Institute of Science, Bangalore, 560 012, India}

\author{A.K. Sood}
\affiliation{Department of Physics, Indian Institute of Science, Bangalore, 560012, India}
\author{Prabal K. Maiti}
\affiliation{Centre For Condensed Matter Theory, Department of Physics, Indian Institute of Science, Bangalore, 560 012, India}

\email{maiti@physics.iisc.ernet.in}
\date{\today}

\begin{abstract}
Experiments and computer simulations demonstrate that water spontaneously fills the hydrophobic 
cavity of a carbon nanotube. To gain a quantitative thermodynamic understanding of this phenomenon, we 
use the recently developed Two Phase Thermodynamics (2PT) method to compute translational and rotational entropies  
of confined water molecules inside single-walled carbon nanotubes and show that the increase
in energy of a water molecule inside the nanotube is compensated by the gain in its rotational entropy. The confined water 
is in equilibrium with the bulk water and the Helmholtz free energy per water molecule of confined water
is the same as that in the bulk within the accuracy of the simulation results. A comparison of 
translational and rotational spectra of
water molecules confined in carbon nanotubes with that of bulk water shows
significant shifts in the positions of the spectral peaks that are directly related to the tube radius. 
\end{abstract}
\maketitle

\section{Introduction}

Understanding the properties of water confined in a one-dimensional channel of diameter comparable to a few molecular 
diameters is important in many technological and biological processes. Such strong confinement leads to
dramatic changes in the structure and dynamics of water. An ideal system for studying such confinement effects 
is water inside single-walled carbon nanotubes (SWCNT). Several simulations and some experiments 
have demonstrated that water spontaneously
fills up the hydrophobic pore of a nanotube when it is immersed in a bath of water \cite{hummer_nature,koga_nature,
majumder_nature,koles_PRL}. Hummer {\it et al.}~\cite{hummer_nature} demonstrated for the first time, via 
classical molecular dynamics (MD) simulations, that narrow (6,6) SWCNTs immersed in a bath of water remain   
filled with water molecules throughout simulation run times of up to 66 ns. The single-file chain of
water molecules exhibits rapid concerted movements along the axis of the tube. 
The hydrogen bond network of water molecules inside a narrow nanotube is very
different from that of bulk water: the number of hydrogen bonded neighbors of a water molecule inside a narrow
nanotube is generally smaller than that in the bulk. This causes the hydrogen bonding energy of water molecules inside a nanotube to be higher than that of water molecules in the bulk.
The increase in the hydrogen bonding energy of water molecules inside a (6,6) nanotube is about 10 kcal/mol, 
of which only about 6 kcal/mol can be recovered from their interactions with the carbon 
atoms. Therefore, considering the energetics, it is surprising that water molecules spontaneously fill up the nanotube.\textcolor{blue}{Although there have been attempts to understand thermodynamics of entry of water molecules in the hydrophobic pore of a narrow nanotube, it is still not well understood. Maibun \textit{et. al.}\cite{chandler_jpcb2003} and Zhou \textit{et. al.}\cite{zhou_jcp2004} have explained filling and emptying transition in hydrophobic channels using ising like lattice-gas model. Vaitheeswaran \textit{et. al.}\cite{hummer_jcp2004} have calculated free energy of transfer for water molecules, from bulk to  carbon nanotube channe for different occupancy of nanotubel, using statistical-mechnical framework for the periodically replicated nanotubes. They have computed free energy of transfer for the water molecules in $13.4\AA$ long nanotube with diameter $8.1\AA$ in virtual equilibrium with bulk water, for the various occupancy  and shown N=6 is most stable filling for this nanotube. In a excellent review rasaiah \textit{et. al.}\cite{hummer_annrev2008} have reviewed thermodynamic, structural and kinetic aspects of water in various kind of confinement including CNT, protein cavity and flurene. Despite all these attempts, fully microscopic description  for this phenomenon is not well understood. }
 This is generally attributed~\cite{hummer_nature} to the rotational entropy gained by the water molecules on entering the tube, a suggestion which has not been verified quantitatively so far. 

\begin{figure}[htbp]
\begin{center}
\includegraphics[scale=.40]{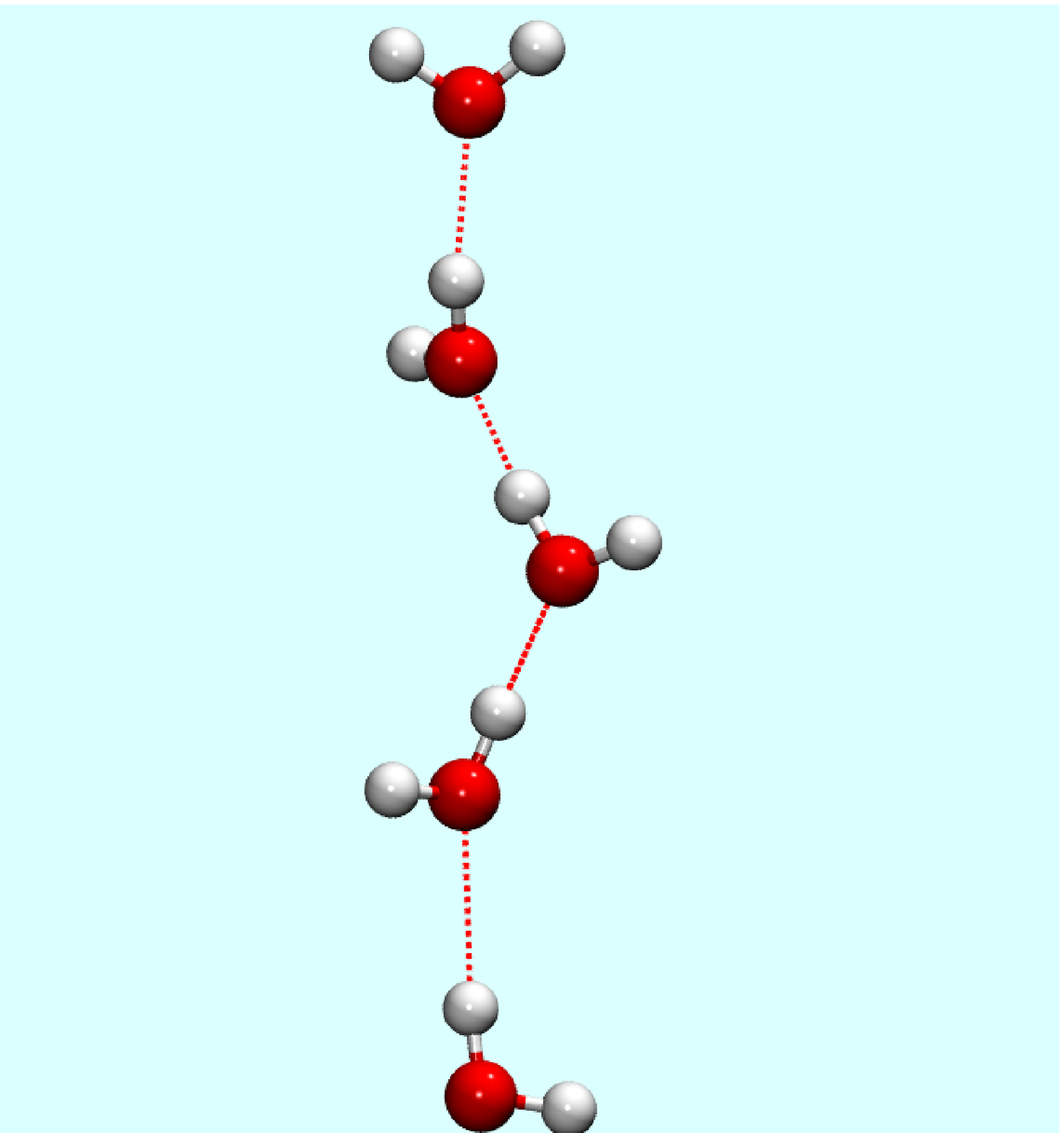} \\
\caption{Single-file arrangement of water molecules inside a (6,6) nanotube.  
One of the hydrogen atoms of each water molecule is not hydrogen-bonded and hence, the corresponding
OH vector can rotate freely.}
\label{water_chain}
\end{center}
\end{figure}

Detailed information about the rotational dynamics of water molecules inside narrow carbon nanotubes is necessary
for understanding the role of rotational entropy in the thermodynamics of entry of water molecules in a nanotube.
Earlier studies~\cite{hummer_pnas,biswa_acs} have shown that water molecules confined 
inside a (6,6) nanotube form an
orientationally ordered chain with the dipole moments of the water molecules pointing either parallel or 
anti-parallel to the axis of the tube (see Fig.\ref{water_chain}). Occasionally, 
there are orientational defects in the chain, where the 
water molecule points perpendicular to the chain. 
As shown in Fig.~\ref{water_chain}, each water molecule in the chain is hydrogen bonded to two water
molecules, one in front and the other behind it. This arrangement of molecules leaves one hydrogen atom for 
each water molecule free which does not participate in any hydrogen bond. The altered hydrogen bond network 
severely modifies the orientational dynamics~\cite{biswa_acs} of the confined water molecules. 
The orientational relaxation 
of the confined water molecules is found to occur in several different time scales and it also exhibits a strong 
directional anisotropy. The slowest relaxation is that of the collective dipole moment (time scale of 
several nanoseconds). This relaxation is mediated by
orientational defects in the hydrogen-bonded chain: a transition from the parallel to the 
anti-parallel (or vice versa) state occurs when a defect is able to hop 
across the whole chain. 
Since the creation of a defect is energetically
unfavorable, collective flips of the dipole moment occur rarely (on a time scale of nanoseconds). 
On the other hand, the relaxation of
the vector joining the hydrogen atoms (HH vector) of each confined water molecule is much faster (with a time scale of 
about 200fs) than that of water molecules in the bulk. 
Maintaining the hydrogen bond network of two hydrogen bonds per water molecule, the HH vector can 
rotate easily about the common dipolar direction in the cylindrically symmetric environment provided by 
the neutral carbon atoms of the nanotube. In addition, the water molecules exhibit another rotational mode,
where there is a fast (time scale of about 60 fs) exchange of the positions of the hydrogen atom participating 
in the hydrogen
bonding in the chain and the one that is free. This exchange is by a non-diffusive angular jump, where the 
mean jump angle is about 50 degrees \cite{biswa_jpcb_2009}
and the mean waiting time between successive jumps is about 1 ps. 
The orientational relaxation of 
water molecules in the bulk is also known to involve large amplitude angular jumps \cite{jump_wat_bulk}.   
In the bulk, the waiting time between two successive jumps is about 2 ps, which is slightly 
reduced in the case of water in nanotubes due to strong confinement. 
Hummer {\it et al.} \cite{hummer_nature} have commented that despite strong hydrophobic 
confinement, the single-file water molecules retain considerable orientational freedom resulting
in a highly degenerate state with a narrow energy distribution. The reorientational processes described
above provide a clear microscopic picture of the ''degenerate states'' involved in the
orientational relaxation. 

In the present study, we investigate how the presence of these microscopic degenerate states affects the 
thermodynamics of water occupancy of the narrow hydrophobic pores of carbon nanotubes.
We present the results of a numerical calculation of the entropy and free energy of water molecules 
using trajectories obtained from atomistic MD simulations. 
For the computation of the entropy of bulk and confined water, we have used the 2PT method of Lin {\it et al.} 
~\cite{Lin_JCP}. 
This is a modification of the quasi-harmonic approximation method where the entropy is determined by treating each 
mode in the vibrational spectrum as a harmonic oscillator. In the present 2PT method, anharmonic effects have been  
explicitly included. 
It has been demonstrated ~\cite{Lin_JCP} that this method yields accurate estimates of thermodynamic quantities from reasonably short MD trajectories (several tens of picoseconds).  
Earlier, the 2PT scheme was successfully used for calculating the entropy of water in different 
domains of PAMAM dendrimers~\cite{Lin_JPCB_2005}, for determining various phases of 
dendrimer liquid crystals~\cite{Lin_JPCB_2004}
and in calculating the relative stability of various forms of aggregates~\cite{Lin_JPCB_2004}.
More recently, it has also been used to compute the entropy of water molecules in major and minor
grooves of DNA~\cite{jana_JPCB}, showing that the entropies are significantly lower than that in bulk water.
In this paper, we compute the entropy of water molecules confined in SWCNTs of various diameters ranging from
6.9 $\AA$ to 10.8 $\AA$ (in standard nomenclature, (5,5) to (8,8)), where water molecules are organized 
in different geometrical arrangement. 
In all cases, we find that the confined water molecules have more rotational entropy than 
the water molecules in the bulk, 
the difference being maximum for molecules inside the narrowest (5,5) nanotubes. The interaction energy of 
each water molecule has also been computed, from which we were able to calculate the average Helmholtz 
free energy ($F=E-TS$). We find that in all cases except (5,5) nanotubes, the free energy of a 
water molecule in the bulk is the same 
as its free energy under confinement, within the
accuracy of our calculation. For water molecules inside a
(5,5) SWCNT (diameter $6.9\AA$) the gain in entropy is not able to compensate the increase in energy. 
This results in reduced occupation of the nanotube by water molecules. 

The rest of the article is arranged as follows. In the next section II, 
we provide the theoretical details of the 2PT method. Section III describes the simulation method and the 
results are described in detail in section IV. Section V contains a summary of out main conclusions.


\section{Two Phase Thermodynamics Method}
The translational density of states (DoS) for solids represents the phonon spectrum~\cite{mcquarrie_book} 
as in the Debye-Einstein 
Model. Thermodynamic quantities for solids can be computed by treating the phonon modes as a system of
non-interacting harmonic oscillators. 
Hard sphere fluids have exponentially decaying velocity auto-correlation function, within low density approximation ~\cite{mcquarrie_book}, hence DoS function can be extracted analytically.
 The 2PT method is based on the premise
that a fluid can be regarded as a two component system consisting of solid and  gas.
Based on this premise, Lin {\it et al.}~\cite{Lin_JCP} suggested that thermodynamic properties of 
fluids can be computed by treating the DoS of fluids as a sum of solid-like [$S^s(\nu)$] and gas-like [$S^g(\nu)$]
contributions. Using this prescription, they have computed very accurate thermodynamic quantities of Lennard-Jones fluids over
a wide range of thermodynamic state points from the DoS function obtained from only 20ps MD trajectory. 
The translational density of states $S(\nu)$ of a system is  defined as the mass-weighted sum of atomic spectral densities
$s^{k}_j(\nu)$,
\begin{eqnarray} \label{dos_def}
S(\nu)=\frac{2}{k_BT}\sum_{j=1}^{N}\sum_{k=1}^{3} m_js^{k}_j(\nu)\,,
\end{eqnarray}
where $m_j$ is the mass of the $j^{th}$ atom, $k$ refers to Cartesian directions $x,y,z$, and
$s^{k}_j(\nu)$ is defined as
\begin{eqnarray}
 s^{k}_j(\nu)=lim_{\tau\to\infty}\frac{|\intop_{-\tau}^{\tau}v^{k}_j{(t)}e^{-i 2\pi \nu t}|^2}{\intop_{-\tau}^{\tau}dt}
\end{eqnarray}
where $v^{k}_j(t)$ is the $k^{th}$ component of the velocity of atom $j$. It can be shown that the atomic spectral density
defined as above can be obtained from the Fourier transform of the velocity 
auto-correlation function (VACF) $c_j^k(t)$\cite{Lin_JCP}:
\begin{eqnarray}
s^{k}_{j}(\nu)=lim_{\tau\rightarrow\infty}\intop_{-\tau}^{\tau}c^{k}_j(t)e^{-i2\pi \nu t}dt\,.
\end{eqnarray}
Therefore, Eq.(\ref{dos_def}) can be rewritten as
\begin{eqnarray}\label{vac2dos}
S(\nu)=\frac{2}{k_{B}T}lim_{\tau\rightarrow\infty}\intop_{-\tau}^{\tau}\sum_{j=1}^{N}\sum_{k=1}^{3}
m_{j}c^{k}_j(t)e^{-i2\pi \nu t}dt\,.
\end{eqnarray}
More generally, one can write
\begin{eqnarray}\label{vac2dos1}
S(\nu)=\frac{2}{k_{B}T} lim_{\tau\rightarrow\infty} \intop_{-\tau}^{\tau}
   C(t) e^{-i 2\pi\nu t}dt
\end{eqnarray}
where $C(t)$ can be either the mass-weighted translational VACF determined from the
center-of-mass velocity ${\bf V}_{i}^{cm}(t)$ of the $i^{th}$ water molecule,
\begin{eqnarray}
C_T(t)=\sum_{i=1}^N<m_i{\bf V}_i^{cm}(t).{\bf V}_i^{cm}(0)>
\end{eqnarray}
or the moment-of-inertia weighted angular velocity auto-correlation function
\begin{eqnarray}
C_R(t)=\sum_{j=1}^3\sum_{i=1}^N<I_{ij}\omega_{ij}(t)\omega_{ij}(0)>
\end{eqnarray}
where $I_{ij}$ and $\omega_{ij}$ are, respectively, the $j^{th}$ components of the moment of inertia tensor and the 
angular velocity of the $i^{th}$ water molecule. Depending on the translational or rotational velocity 
correlation used in Eq.(\ref{vac2dos1}), one obtains the translational or rotational DoS. 

In the 2PT method, the DoS
is decomposed into a solid-like non-diffusive component and a gas-like diffusive 
component, $S(\nu)=S^g(\nu)+S^s(\nu)$, using the fluidity factor $f$ which is a measure of the fluidity of system.
The value of $f$ is computed in terms of the dimensionless diffusivity 
$\Delta$ using the {\it universal equation}~\cite{Lin_JCP} 
\begin{eqnarray}\label{f_eqn}
2 \Delta^{-9/2}f^{15/2}-6\Delta^{-3}f^{5}-\Delta^{-3/2}f^{7/2}+6\Delta^{-3/2}f^{5/2}+2f-2=0.
\end{eqnarray}
The diffusivity $\Delta$ can be uniquely determined for a thermodynamic state of the system using following
equation: 
\begin{eqnarray}\label{delta_eqn}
\Delta(T,\rho,m,S_0)=\frac{2S_0}{9N}\left(\frac{6}{\pi}\right)^{2/3}\left(\frac{\pi k T}{m}\right)\rho^{1/3},
\end{eqnarray}
where $S_0=S(0)$ is the zero-frequency component of the DoS function (translational or rotational).
Knowing $f$ from Eq.(\ref{f_eqn}) and Eq.(\ref{delta_eqn}), the gas-like diffusive  component of the DoS can be obtained using a hard-sphere diffusive
model:
\begin{eqnarray}\label{hs_dos}
S^{g}(\nu)=\frac{S_0}{1+[\frac{\pi s_0 \nu}{6fN}]^2}.
\end{eqnarray}

Lin {\it et al.}~\cite{Lin_JCP} used such a decomposition only for the translational
DoS of monatomic fluids. In a recent development, Lin {\it et al. }~\cite{lin_maiti_new} have shown that
for polyatomic fluids, improved estimates of the rotational entropy can be obtained if such a gas-solid decomposition is 
used for the rotational DoS as well. In this case, separate fluidity factors $f$ are determined for the
translational and rotational DoS using rotational and translational diffusivities in Eq.(\ref{f_eqn})
and the gas-like component of entropy is determined by Eq.(\ref{hs_dos}) with $S_0$ being $S_{tran}(0)$ or 
$S_{rot}(0)$ for translational  and rotational cases, respectively. Given such a DoS 
decomposition, every thermodynamic quantity has contributions from solid-like and gas-like
DoS functions with appropriate weight functions:
\begin{eqnarray}\label{thermodynamics}
A_m=\int_{0}^{\infty}d\nu S_{m}^{g}(\nu)W_{A,m}^{g}+
\int_{0}^{\infty}d\nu S_{m}^{s}(\nu)W_{A,m}^{s}\,,
\end{eqnarray}
where $m$ can be translational, rotational or vibrational. For the rigid water model used in our calculation, 
there is no contribution from intra-molecular vibrations. Details of the
weight functions can be found in Ref.~\cite{lin_maiti_new}.
Different water models have been found to yield very accurate results for thermodynamic quantities over a large range
of thermodynamics state points using this improved 2PT scheme.

\section{Simulation}
We have performed a series of MD simulations of open-ended armchair SWCNTs immersed in a bath of water. 
We have used nanotubes of four different diameters, (5,5), (6,6), (7,7) and (8,8) having 
diameters of $6.8\AA$, $8.2\AA$, $9.6\AA$ and $10.9\AA$, respectively. 
Interactions between various atoms were modeled  by classical force-fields; carbon atoms were modelled as  
chargeless Lennard-Jones particles with $sp^{2}$ hybridization (AMBER03 atom type ``CA'') with 
parameter values given in our earlier paper~\cite{biswa_JCP}. The equilibrium C-C bond length and the C-C-C angle 
were taken to be $1.44\AA$ 
and $2\pi/3$ respectively, with spring constants of $938 Kcal/mol/{\AA}^2$ and $126Kcal/mol/{rad}^2$, respectively. 
Water was modeled as a rigid molecule with the TIP3P  potential. 
Electrostatic interactions were computed  using the Particle Mesh Ewald (PME) method with a real space cutoff
of $10.5\AA$ and cubic B-spline interpolation. The same cut-off was used for Lennard-Jones interactions
with a neighbor-list update frequency of 10. Each carbon nanotube had a length of $56\AA$ with its axis 
fixed along $z$-axis during the whole simulation. The simulation unit box was created by solvating the carbon nanotube with 
$15\AA$ water solvation shells in all three directions. Before using the MD trajectory for analysis, 
an energy minimization was done to remove bad contacts, followed by slow heating for 100 ps to reach the 
system temperature of 300K. The simulation box was then equilibrated in the NPT
($P=1$ atm,$T=300$K) ensemble for at least 1 ns to get the correct density. Berendsen weak coupling method was used 
to maintain constant temperature and pressure with coupling constants of 0.5 ps and 0.2 ps, respectively, for
temperature and pressure bath coupling. Finally, simulations were performed in the NVT ensemble with integration 
time-step of 1.0 fs. To compute the VACF, coordinates and velocity components of each atom were stored for 40 ps with 4 fs saving 
frequency. To get better statistical data, four independent trajectories were used for further computations. 
All the simulations reported here were done using the AMBER10~\cite{amber} MD simulation package and
visualization of simulation trajectories was done using VMD~\cite{vmd}.

To compute the entropy of bulk water, 100 water molecules, which remain at least  $10\AA$ away
from the nanotube, were chosen at random. For confined water, only those water molecules which  stay inside the tube during
the complete production run were selected. Translational and rotational DoS were computed 
by decomposing the velocity into translational and rotational components at every step, 
computing velocity auto-correlation functions and taking their 
Fourier transform. The diffusivity of the system was computed using Eq.(\ref{delta_eqn}).
This diffusivity was used to compute the fluidity factor $f$ using Eq.(\ref{f_eqn}). Knowing the fluidity factor $f$,
the gas phase DoS was computed using Eq.(\ref{hs_dos}) and the solid phase DoS was determined by subtracting the 
gas phase DoS from the total DoS. Note that such gas-solid decomposition was done for both translation and
rotational DoS with fluidity factor computed separately for each case. The computed fluidity factors are given in Table I. 
Once we have this DoS decomposition, the entropy was computed using Eq.(\ref{thermodynamics}). 
The energy of each water molecule was computed by decomposing the potential energy per atom using MPSIM~\cite{mpsim}. 
This decomposition was done by assigning equal values of energy to both atoms in the case of pair-wise 
interaction terms like the VDW energy and the bond energy. 
 Long range Coulomb interaction energy of each atom was computed by observing 
that various terms in the Ewald energy are expressed in terms of the electrostatic potential at charge sites. 
Knowing the electrostatic potential at every charge site, the interaction energy of the $i^{th}$ atom is the 
charge $q_i$ multiplied by the potential at the site of the atom.

\section{Results}
\begin{figure}[htbp]
\begin{center}
\includegraphics[scale=.9]{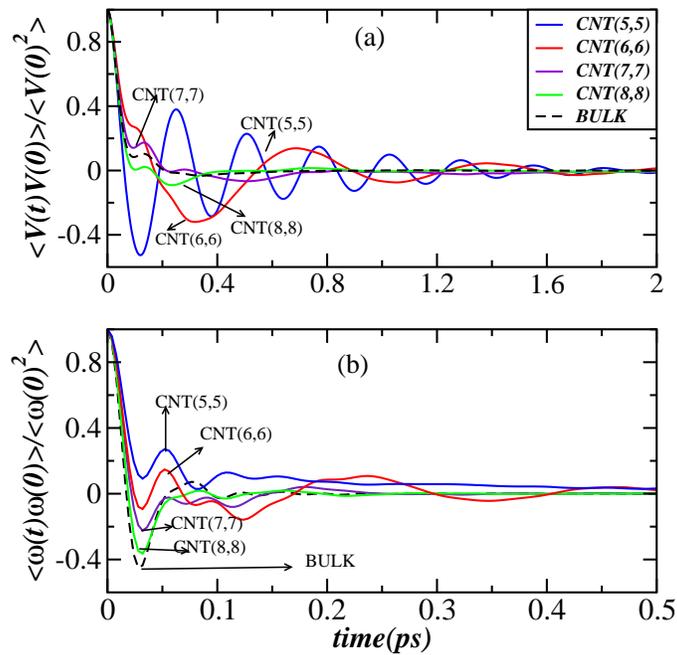} \\
\caption{Translational (a) and rotational (b) velocity
auto-correlation functions of water molecules in the bulk and of those
confined inside (8,8), (7,7), (6,6), and (5,5) nanotubes of length $54\AA$, calculated at T = 300K.}
\label{vac_tip3p}
\end{center}
\end{figure}

\begin{figure}[htbp]
\begin{center}
\includegraphics[scale=.9]{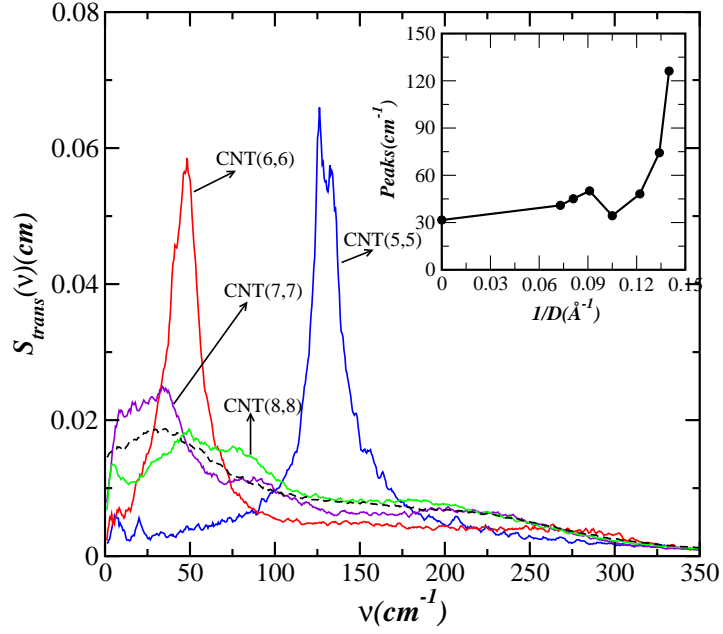} \\
\caption{Translational DoS function of water molecules in the bulk and of those
confined inside (8,8), (7,7), (6,6) and (5,5) nanotubes of length $54\AA$, calculated at T = 300K. Note the gradual shift in the peak frequencies as the diameter of the tube is increased. The inset shows the variation of the peak position with the inverse of the tube diameter.}
\label{pwr_trans}
\end{center}
\end{figure}

Translational and rotational VACFs 
are shown in Fig.~\ref{vac_tip3p} for water molecules in the bulk, as well as for those 
confined in (8,8), (7,7), (6,6) and (5,5) nanotubes. Fig.~\ref{pwr_trans} and Fig.~\ref{pwr_rot} show translational and rotational 
power spectra, respectively, for the same cases. Both translational and rotational VACFs 
in Fig.~\ref{vac_tip3p} show high-frequency oscillations for the water confined in narrow (5,5) and (6,6) nanotubes. 
These oscillations are more pronounced for (5,5) than for (6,6). The VACFs 
for water molecules confined in (7,7) and (8,8) nanotubes are similar to those of bulk water. 
Prominent peak appears in the translational power spectrum at $121 cm^{-1}$ for (5,5) 
and at $47 cm^{-1}$ for (6,6) nanotubes. For  wider nanotubes, (7,7) and (8,8), where water molecules are not 
restricted to be arranged in single-file manner, the power spectrum (DoS function) is similar in shape to that of bulk water. 
We have observed a systematic variation of the position of this low-frequency peak with the diameter of the nanotube, 
as shown in Fig.~\ref{pwr_trans}. 

\begin{figure}[htbp]
\begin{center}
\includegraphics[scale=.9]{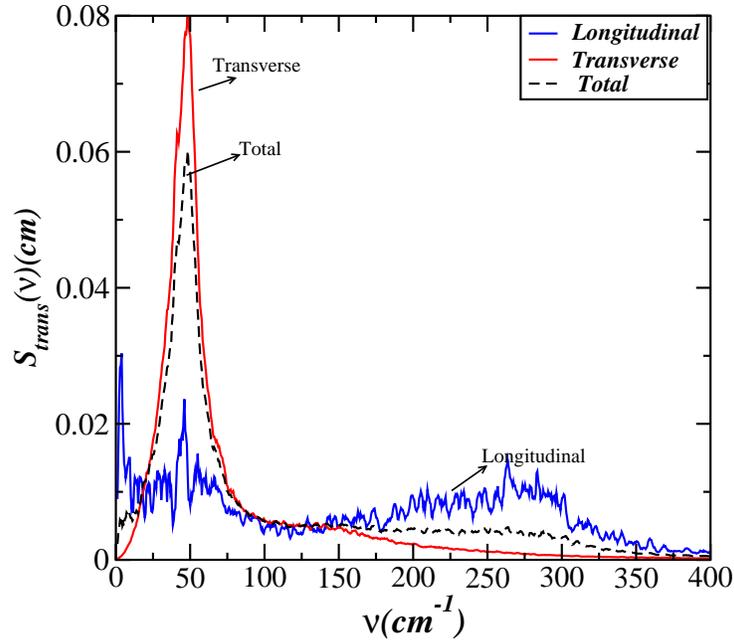} \\
\caption{Decomposition of the translational DoS function in longitudinal and transverse directions for water in a (6,6) nanotube of length $54\AA$, calculated at T = 300K. It is obvious that the peaks in the DoS arise from transverse motion.}
\label{pwr_decomp_66}
\end{center}
\end{figure}

In the case of bulk water, the low frequency
peak around $32cm^{-1}$ has been attributed to hindered translation  in a local structural cage~\cite{marti_jcp_comm}.
To find the molecular origin of the prominent peaks appearing in the DoS function of (5,5) and (6,6) nanotubes,
we decomposed the translational velocity along longitudinal (parallel to the axis of the nanotube) and transverse 
(perpendicular to the nanotube axis) components and determined the DoS for motion in these two directions.
Results shown in Fig.~\ref{pwr_decomp_66} suggest that the peak in Fig.~\ref{pwr_trans} appears 
due to vibrational motion in direction perpendicular to  axis of the nanotube.  
The confining potential is  stronger in the case of the narrower (5,5)
tube and hence the peak is blue shifted. For (7,7) and (8,8) nanotubes, water molecules have bulk-like environments and
hence, they have DoS similar to that of bulk water with slight blue shift.
Positions of these DOS peaks for translational 
modes have systematic variation with tube diameter as shown in the 
inset of  Fig.~\ref{pwr_trans}. Similar behavior is also 
observed for the peak positions of the rotational DOS as shown in the inset of  Fig.~\ref{pwr_rot}.

Due to single-file arrangement of water molecules inside (5,5) and (6,6) nanotubes,
they have more freedom for rotational motion, and peaks in the rotational spectrum shift to 
lower frequencies as shown in Fig.~\ref{pwr_rot}. A distinct peak in the rotational DoS at $ 146cm^{-1}$ is observed 
for water molecules confined in (6,6) nanotubes, 
whereas for the (5,5) nanotube, this peak appears to have merged with the zero-frequency diffusive peak. 
The emergence of these peaks is a result of the change in the librational motion of water molecules when they are confined in narrow nanotubes. 
The reduction in the average number of hydrogen bonds per water molecule from $4$ in the bulk to $2$ for single-file water in narrow nanotubes leads to less constrained rotation and faster rotational diffusion of water molecules inside these nanotubes.

Table I gives a summary of the results of our entropy calculations. For bulk water, the calculated total 
(translational plus rotational) entropy
is 4.95 Kcal/mol for $T=300$K which is quite close to the experimental value of 5.01 Kcal/mol~\cite{nist_data} 
under the same conditions. This value is also close to the results obtained from a recently 
developed methods based on cell theory~\cite{nist_data} which also uses the harmonic approximation for 
calculating the entropy, as in the 2PT method. This gives us confidence that the rotational entropy of 
water molecules obtained from the 2PT method is accurate enough to explain the thermodynamics of entry of water inside the hydrophobic pores of nanotubes. \textcolor {red}{Here we would like to clarify that the number of water molecules inside the nanotube is not constant during the simulation. Hence the values of $<E> $and $<S>$ reported here correspond to averages in the grand canonical ensemble. Therefore, $A=<E>-T<S>$ is not the canonical Helmholtz free energy, here onwards we represents $<E>$ , $<S>$ ,$A$ by $<E>^{GC}$ , $<S>^{GC}$ and $A^{GC}$ to indicate that these values belong to grand canonical ensemble for confined water. Note that this difference between the this free energy and the conventional canonical Helmholtz free energy will be  of the negligible when the number of particles in the nanotube is very large\cite{largeN_approx}. This however is not the case as the number of particles inside nanotube in our case is around 20. It may be pertinent to remark that the calculation of Helmholtz free energy for the water molecules inside the nanotube is done by Vaitheeswaran {\it et. al.}\cite{hummer_jcp2004} is  for a fixed number of water molecules inside the nanotube,  hence the relation used  there(Eq. 16 of ref. ~6 ) is not valid in our case. }

From Table I, we see that for water inside nanotubes, when the diameter
of the nanotube is so small that there is 
only a single chain of water molecules inside the nanotube (i.e. for (5,5) and (6,6) nanotubes), the rotational entropy
is substantially higher in comparison to the value of 0.97 kcal/mol for bulk water.  
There is a 70-120\% increase in the rotational component of the entropy for the confined water molecules
compared to that in the bulk. This increase in rotational entropy is due to unhindered rotation of the dipole vector in the absence of a three-dimensional network of hydrogen bonds that is present in the bulk.  
Due to single file arrangement of water inside nanotubes, each water molecule can rotate freely in such a way 
that the component of the dipole moment of every confined water molecule along the nanotube axis remains 
nearly constant, while the azimuthal component relaxes very fast, providing more rotational freedom. 
The resulting gain in rotational entropy helps to compensate 
the energy cost due to the loss of hydrogen bonds inside a nanotube, as well as the loss in translational entropy. 
Because of this compensation, the  free energy $A^{GC}$ per water molecule inside (6,6), (7,7), and (8,8) nanotubes 
is the same as that for bulk water within the accuracy of our calculation. \textcolor {blue}{ Vaitheeswaran {\it et. al.}\cite{hummer_jcp2004} have computed free energy of transfer and found similar entropy  gain for water molecules inside nanotube for occupancy of 5 for $14\AA$ long tube but they have observed that it is energetically favourable for water molecules to be inside hydrophobic channel which is quite against  to the  intuition. Change  in energy is correctly captured in our calculation which reveals energy loss inside nanotube.}

\begin{figure}[htbp]
\begin{center}
\includegraphics[scale=0.9]{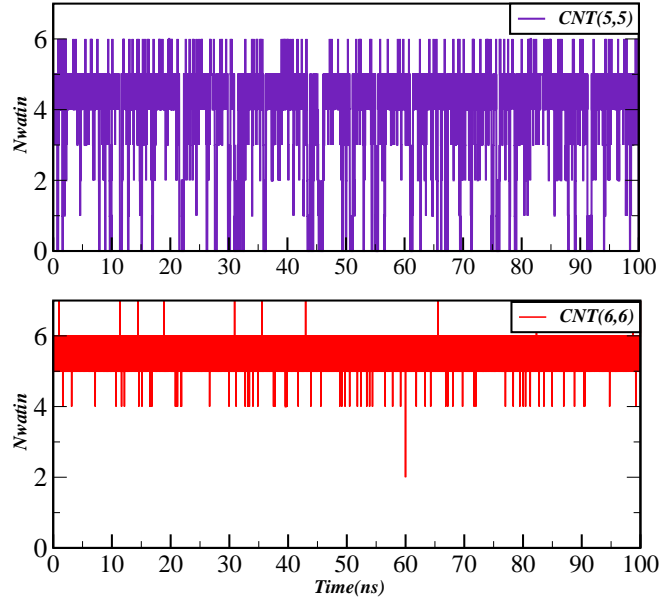} \\
\caption{Number of water molecules inside 13.4 $\AA$ long (5,5) (top panel) and (6,6) (bottom panel) nanotubes as function of time at T = 300K. During 100 ns of simulation time, the (5,5) nanotube have many instances when 
there is no water inside the tube, while inside the (6,6) nanotube, at least four water molecules are always 
present (except once). This is consistent with the computed free energy values.}
\label{nwatin_time}
\end{center}
\end{figure}

\begin{figure}[htbp]
\begin{center}
\includegraphics[scale=0.9]{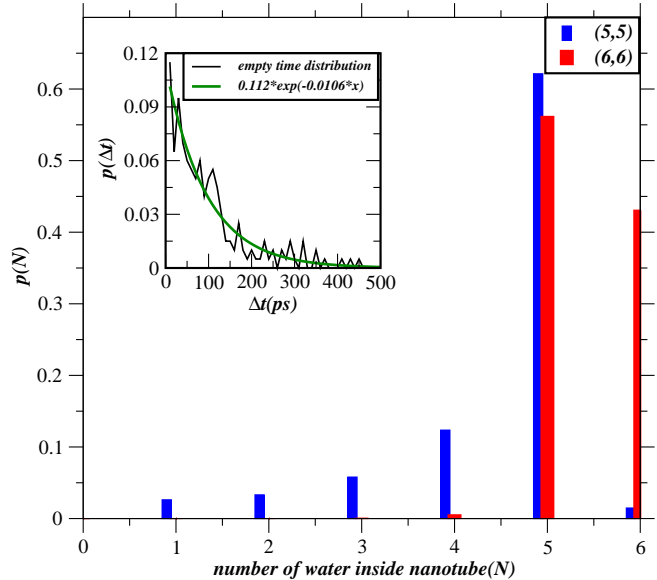} \\
\caption{Probability of having a particular  number of water molecules inside 13.4 $\AA$ long (5,5) and (6,6) nanotubes at T = 300K. 
 The inset shows the probability distribution of time intervals during which the (5,5) nanotube remains empty, fitted to an exponential function.}

\label{watin_histo}
\end{center}
\end{figure}

The Helmholtz free energy per water molecule inside a (5,5) nanotube is 
around 1 kcal/mol higher then that in bulk water. 
In spite of this higher free energy, water spontaneously 
enters inside (5,5) nanotube. To have a microscopic understanding of this phenomenon,  
we have performed  very long simulations (total duration of 100 ns)  for  (5,5) and (6,6) nanotubes 
of length $13.4~\AA$ immersed in water
and found that for the (5,5) nanotube, there are many emptying transitions when there are no
water molecules inside the nanotube. In fact, for the 100 ns long simulation, the nanotube remains empty 
for almost 13 ns, which is  13\% of 
the simulation time (see Fig.~\ref{nwatin_time}). In contrast, for the (6,6) nanotube, there are always 
at least 2 water molecules inside the 
nanotube in a 100 ns simulation run. In Fig.~\ref{watin_histo}, we plot the occupation probability of 
water molecules
for (5,5) and (6,6) nanotubes. In the case of the (6,6) nanotube, the occupancy is peaked at 5 water 
molecules (see Fig.~\ref{watin_histo}) while the (5,5) nanotube has 
considerable weights for smaller numbers of water molecules due to the occurrence of many emptying transitions. 
This is consistent with the results of our free energy calculation. Water molecules may enter a (5,5) nanotube 
due to thermal fluctuations even if the free energy
of a water molecule inside the nanotube is higher than that in the bulk.
However, the higher free energy of a water molecule inside a (5,5) nanotube results in a
reduction in the average occupation of the nanotube by water molecules in long simulations. 
Interestingly, we find that the probability of the durations during which the (5,5) nanotube 
remains empty can 
be fitted to an exponential distribution, as shown in the
inset of Fig.~\ref{watin_histo}.

To check the presence of finite-size effects in our entropy calculations, we 
have carried out these computations for three different lengths of tubes (20, 40, and 60 unit cells) and found 
the same value of the entropy in all three cases. 

\section{Conclusion}
We have shown that water molecules have significant gain in rotational entropy when they
come inside narrow carbon nanotubes. This gain is considerable for the narrower tubes for which water molecules  
inside the tube form single-file chains. For such tubes, the gain in rotational entropy compensates the increase in energy
due to a reduction in the number of hydrogen bonds per water molecule, so that the average free energy of a water
molecule inside the tube is the same (except for the narrowest (5,5) tubes) as that in the bulk. This explains the
observation that water goes inside narrow carbon nanotubes in spite of the hydrophobic nature of the cavity. 
As the diameter of the tube increases, entropy and energy values approach 
those of bulk water. We have also observed considerable shifts in the peaks of the
DoS functions which might serve as spectroscopic signatures of the confined water. 
\section{Acknowledgment} We thank DST, India for financial support. 
HK would like to thank University Grants Commission, India for Junior Research Fellowship.

\begin{figure}[htbp]
\begin{center}
\includegraphics[scale=.9]{figure4.eps} \\
\caption{Comparison of rotational 
DoS of water molecules in the bulk and of those confined inside (8,8), (7,7), (6,6), and (5,5) nanotubes of length $54\AA$, calculated at T = 300K. 
The inset shows the dependence of the peak positions
on the inverse of the tube diameter. Triangles $(\triangleright)$ denote the positions of the 
first peak and circles $(\bullet)$ denote the positions of the second peak of the 
rotational DoS.}
\label{pwr_rot}
\end{center}
\end{figure}


\begin{table}
\label{free_energy_table}
\begin{ruledtabular}
\begin{tabular}{lllllll}

$ $ & $<E>^{GC} \footnotemark[1]$ & $T<S_{trans}>^{GC}$ & $T<S_{rot}>^{GC}$ & $A^{GC}$ &$f_{trans}\footnotemark[2]$& $f_{rot}$\footnotemark[2]  \\ \hline 
$(5,5)$ & $ -8.26 \pm (0.20)$ & $2.97\pm(.20)$ & $2.16 \pm (0.18)$ & $-13.39 \pm (0.58)$& $0.12\pm(.04) $ & $0.42\pm(.03) $ \\ 
$(6,6)$ & $ -8.89 \pm (0.20)$ & $3.84\pm(.20)$ & $1.67 \pm (0.15)$ & $-14.40 \pm (0.55)$& $0.15\pm(.04) $ & $0.30\pm(.03) $\\ 
$(7,7)$ & $ -8.67 \pm (0.20)$ & $4.04\pm(.10)$ & $1.42 \pm (0.07)$ & $-14.13 \pm (0.37)$& $0.17\pm(.01) $ & $0.15\pm(.03) $\\ 
$(8,8)$ & $ -9.38 \pm (0.10)$ & $3.67\pm(.10)$ & $1.12 \pm (0.07)$ & $-14.17 \pm (0.27)$& $0.20\pm(.02) $ & $0.10\pm(.01) $\\ 
$BULK $ & $ -9.45 \pm (0.02)$ & $4.02\pm(.05)$ & $0.93 \pm (0.01)$ & $-14.40 \pm (0.08)$& $0.34\pm(0.005) $ & $0.08\pm(.004) $\\ 
\end{tabular}
\end{ruledtabular}
\footnotetext[1]{In Kcal/mol}
\footnotetext[2]{Fluidity factor}
\footnotetext[3]{Superscript GC indicates that values quoted here belongs to Grand canonical ensemble for confined water molecules. }
\caption{The energy, entropy and free energy per water molecule($A^{GC}=<E>^{GC}-T<S>^{GC}$) in Kcal/mol and fluidity factors at T = 300K for the water molecules confined in $54\AA$ long tubes of various diameters and for bulk water calculated in NVT ensemble.}
\end{table}

\end{document}